\documentstyle[twoside,fleqn,espcrc2]{article}

\title{The $q \bar{q}$ semirelativistic interaction in the Wilson loop 
       approach}
\author{N. Brambilla$^{\rm a~*}$ and A. Vairo 
\address{Institut f\"ur Theoretische Physik, Universit\"at Heidelberg \\
         Philosophenweg 16, D-69120 Heidelberg, FRG}
\thanks{Alexander von Humboldt Fellow }}

\begin{document}

\begin{abstract}
The complete $q \bar{q} $ semirelativistic interaction is obtained 
as a gauge-invariant function of the Wilson loop and 
its functional derivatives. The approach is suitable for analytic
evaluations as well as for lattice calculations. Here we consider 
three different models for the Wilson loop non-perturbative behaviour
and discuss the related semirelativistic dynamics. 
\end{abstract}

\maketitle

\section{INTRODUCTION}

The derivation of the $q \bar{q} $ potential is a longstanding problem;
in Ref. \cite{Wilson,eichten} the static and the spin-dependent potentials
were first obtained. Recently, working with a path integral 
representation for the quark propagator in the external field, the 
complete $1/ m^2$ $q\bar{q}$ potential was obtained as a 
gauge-invariant function of the Wilson loop and its functional derivatives 
\cite{BCP}:  
\begin{eqnarray}
&~&\int_{t_{\rm i}}^{t_{\rm f}} dt V_{{\rm q} \bar{{\rm q}}} = 
i \log \langle W(\Gamma) \rangle 
\nonumber\\
&~&\!\!\!\!\!\!\!\!\!\!\!
- \sum_{j=1}^2 \frac{g}{m_j} \int_{{\Gamma}_j}dx^{\mu} 
\left( S_j^l \, 
\langle\!\langle \hat{F}_{l\mu}(x) \rangle\!\rangle \right.
 \nonumber \\
&~&\!\!\!\!\!\!\!\!\!\!\!
-\frac{1}{2m_j} S_j^l \varepsilon^{lkr} p_j^k \, 
\langle\!\langle F_{\mu r}(x) \rangle\!\rangle 
- \left. \frac{1}{8m_j} \, 
\langle\!\langle D^{\nu} F_{\nu\mu}(x) \rangle\!\rangle  \right)
\nonumber \\
&~&\!\!\!\!\!\!\!\!\!\!\!
- \frac{1}{2} \sum_{j,j^{\prime}=1}^2 \frac{ig^2}{m_jm_{j^{\prime}}}
{\rm T_s} \int_{{\Gamma}_j} dx^{\mu} \, \int_{{\Gamma}_{j^{\prime}}} 
dx^{\prime\sigma} \, S_j^l \, S_{j^{\prime}}^k
\nonumber\\
&~& \!\!\!\!\!\!\!\!\!\!\!
\times \left( \, 
\langle\!\langle \hat{F}_{l \mu}(x) \hat{F}_{k \sigma}(x^{\prime})
\rangle\!\rangle  - \, 
\langle\!\langle \hat{F}_{l \mu}(x) \rangle\!\rangle
\, \langle\!\langle \hat{F}_{k \sigma}(x^{\prime}) \rangle\!\rangle  \right) 
\label{potential}
\end{eqnarray}
where
\begin{equation}
W(\Gamma) \equiv  
{\rm P\>} \exp \left[i g \oint_{\Gamma} dx^\mu A_\mu (x) \right] ,
\label{wgamma}
\end{equation}
$\langle ~~ \rangle$ is the normalized average over the gauge fields 
$A_\mu$ and 
$$
\langle\!\langle f(A) \rangle\!\rangle \equiv
\langle f(A) W(\Gamma) \rangle / \langle f(A) W(\Gamma) \rangle.  
$$

The closed loop $\Gamma$ is defined by the quark (anti-quark) 
trajectories ${\bf z}_1 (t)$ (${\bf z}_2 (t)$) running from ${\bf y}_1$ 
to ${\bf x}_1$ (${\bf x}_2$ to ${\bf y}_2$) along with 
two straight-lines at fixed time connecting ${\bf y}_1$ to ${\bf y}_2$ 
and ${\bf x}_1$ to ${\bf x}_2$. Moreover we have 
\begin{equation}
g \langle\!\langle F_{\mu\nu}(z_j) \rangle\!\rangle = (-1)^{j+1}
{\delta i \log \langle W(\Gamma) \rangle \over \delta S^{\mu\nu} (z_j)}  ,
\label{e20}
\end{equation}
\begin{eqnarray}
&~& \!\!\!\!\!\!\!\!
g^2 \left(\langle\!\langle F_{\mu\nu}(z_1) 
F_{\lambda\rho}(z_2) \rangle\!\rangle 
- \langle\!\langle F_{\mu\nu}(z_1) \rangle\!\rangle 
  \langle\!\langle F_{\lambda\rho}(z_2) \rangle\!\rangle \right)
\nonumber \\
&~& = - i g {\delta\over \delta S^{\lambda\rho}(z_2)} 
\langle\!\langle F_{\mu\nu}(z_1) \rangle\!\rangle,
\label{e21}
\end{eqnarray}
where  $\delta S^{\mu\nu} (z_j) =  dz_j^\mu \delta z_j^\nu 
- dz_j^\nu \delta z_j^\mu$. As a consequence we have that the actual 
form of the potential in (\ref{potential}) is completely known once 
the Wilson loop behaviour is given.

The semirelativistic $q\bar{q}$ potential can be written as \cite{BCP,BMP}
\begin{equation} 
V_{{\rm q} \bar {\rm q}} = V_0 + V_{\rm VD} + V_{\rm SD} ,
\end{equation}
with
\begin{eqnarray}
&& 
i \log \langle W(\Gamma) \rangle =\int_{t_{i}}^{t_{f}} dt ~
V_0 (r(t)) + V_{\rm VD}({\bf r}(t)) \, ,
\nonumber \\
&&\!\!\!\! 
V_{\rm VD}({\bf r}(t)) =  {1\over m_1 m_2} 
\bigg\{ {\bf p}_1\cdot{\bf p}_2 V_{\rm b}(r) 
\nonumber\\
&&\!\!\!\! 
+ \left( {1\over 3} {\bf p}_1\cdot{\bf p}_2 - 
{{\bf p}_1\cdot {\bf r} \>~ {\bf p}_2 \cdot {\bf r} \over r^2}\right) 
V_{\rm c}(r) \bigg\}_{\rm W}
\nonumber \\
&&\!\!\!\! 
+ \sum_{j=1}^2 {1\over m_j^2} \bigg\{ p^2_j V_{\rm d}(r) 
\nonumber \\
&&\!\!\!\! 
+ \left( {1\over 3} p^2_j - 
{{\bf p}_j\cdot {\bf r} \>~ {\bf p}_j \cdot {\bf r} \over r^2}\right) 
V_{\rm e}(r) \bigg\}_{\rm W} , 
\label{vd}
\end{eqnarray}
where ${\bf r}(t) \equiv {\bf z}_1(t) - {\bf z}_2 (t)$ and 
the symbol $\{\> ~\}_{\rm W}$ stands for the Weyl ordering 
prescription.
The spin dependent interaction appearing in (\ref{potential}) 
has a transparent physical meaning the first term being the magnetic 
interaction, the second the Thomas precession, the third the Darwin term
and the last one the spin-spin interaction. However, it can be rewritten
in the usual Eichten--Feinberg form \cite{BV}:
\begin{eqnarray}
& & V_{\rm SD} =
{1\over 8} \left( {1\over m_1^2} + {1\over m_2^2} \right) 
\Delta \left[ V_0(r) +V_{\rm a}(r) \right] 
\nonumber\\
& &  
\!\!\!\!\!\!\!\!\!\!
+ \sum_{j=1,2}
 \left( {(-1)^{j+1}\over 2 m_j^2} {\bf L}_j \cdot {\bf S}_j 
      \right) 
       {1\over r}  {d \over dr} \left[ V_0(r)+ 2 V_1(r) \right]
\nonumber \\
& & 
\!\!\!\!\!\!\!\!\!\!
+ {1\over m_1 m_2}
\left( {\bf L}_1 \cdot {\bf S}_2 - {\bf L}_2 \cdot {\bf S}_1 \right) 
{1\over r} {d \over dr} V_2(r) 
\nonumber \\
& &
\!\!\!\!\!\!\!\!\!\!
+ {1\over m_1 m_2}  
\left( { {\bf S}_1\cdot{\bf r} \> {\bf S}_2\cdot{\bf r}\over r^2} 
- {1\over 3} {\bf S}_1 \cdot {\bf S}_2 \right) V_3(r) 
\nonumber \\
&& 
\!\!\!\!\!\!\!\!\!\!
+ {1\over 3 m_1 m_2} 
{\bf S}_1 \cdot {\bf S}_2  V_4(r). 
\label{sd}
\end{eqnarray}
It is clear that all the dynamics is contained in the $V_i (r)$ 
of Eqs.(\ref{vd})-(\ref{sd}) and these are unambiguous functions
of the Wilson loop (cf. (\ref{potential})).
 Similar expressions for the $V_i$ in terms of insertions
on the static Wilson loop were obtained in \cite{BMP}. These 
expressions are suitable for lattice evaluation and were used 
in \cite{bali} to obtain lattice predictions for the semirelativistic 
interaction. However, this is also the ideal framework 
to formulate hypothesis on the Wilson loop behaviour 
(and so on the confinement mechanism) to be checked on the 
lattice and on the experimental data. In the following we
use three different models for the QCD vacuum to obtain an analytic
behaviour of the Wilson loop in the non-perturbative region and 
to predict the semirelativistic quark dynamics. Comparison among the results
allows us to get some insight in the mechanism of confinement \cite{BV}.

\section{THE CONFINING SEMIRELATIVISTIC DYNAMICS}

\subsection{Minimal area law model (MAL)}

In Ref. \cite{BCP,BV} $\langle W(\Gamma) \rangle$ 
was approximated by the sum of a perturbative part given at the 
leading order by the gluon propagator $D_{\mu\nu}$ 
and a non-perturbative part given by the value of the minimal area of 
the deformed Wilson loop of fixed contour $\Gamma$ plus a perimeter 
contribution $\cal P$:
\begin{eqnarray}
& & i \log \langle W (\Gamma) \rangle =  
 - \frac{4}{3} g^2 \oint_{\Gamma}dx^{\mu}_1 \oint_{\Gamma} dx^{\nu}_2 
\nonumber \\
& & \quad 
\times iD_{\mu \nu} (x_1-x_2) + \sigma S_{\rm min} + {C\over 2} {\cal P}.
\label{mal}
\end{eqnarray}

Denoting by $u^{\mu}=u^{\mu}(s,t)$ the equation of 
any surface with contour $\Gamma$ ($s \in [0,1],\,
t \in [t_{\rm i},t_{\rm f}], \, 
u^0(s,t)=t, \, {\bf u}(1,t)= {\bf z}_1(t), 
\, {\bf u}(0,t)= {\bf z}_2(t) \,$)
we have:
\begin{eqnarray}
& &  S_{\rm min}  =  {\rm min} \int_{t_{\rm i}}^{t_{\rm f}}
dt \, \int_0^1  ds
\nonumber\\
& & 
\!\!\!\!\!\!\!\!
\times \left[
\left( \frac{\partial u^{\mu}}{\partial t} \frac{\partial u_{\mu}}
{\partial s} \right)^2 
- \left( \frac{\partial u^{\mu}}{\partial t} \frac{\partial u_{\mu}}
{\partial t} \right) \left( \frac{\partial u^{\mu}}{\partial s}
\frac{\partial u_{\mu}}{\partial s} \right)
\right]^{\frac{1}{2}}
\nonumber\label{smin}
\end{eqnarray}
which coincides with the Nambu--Goto action.
Up to the order $1/m^2$ the minimal surface  can be identified 
exactly  with the surface spanned by the 
straight-line joining $(t,{\bf z}_1(t))$ to $(t,{\bf z}_2(t))$.
From Eq. (\ref{mal}) $V_{{\rm q}\bar{{\rm q}}}$ is obtained \cite{BCP,BMP}. 
In particular  for the static potential we have 
\begin{equation}
V_0 = -{4\over 3} {\alpha_s \over r} + \sigma r + C. 
\end{equation}
See Tab. 1 for the complete results.

\subsection{Stochastic vacuum model}

Using the non-Abelian Stokes theorem and the cumulant expansion 
it is possible to write
\begin{eqnarray}
& & \langle W(\Gamma) \rangle 
\nonumber\\
& &\!\!\!\!\!\!\!
= \left\langle {\rm P} \>
\exp \left( ig \int_S dS_{\mu\nu}(u) F_{\mu\nu}(u,x_0) \right) \right\rangle 
\nonumber \\ 
& &\!\!\!\!\!\!\! = 
 \exp \sum_{j=1}^{\infty} {(ig)^j\over j!} 
\int_S dS_{\mu_1\nu_1}(u_1) \cdots \int_S dS_{\mu_j\nu_j} (u_j) 
\nonumber\\
& & 
\times  \langle F_{\mu_1\nu_1}(u_1,x_0) \dots F_{\mu_j\nu_j}(u_j,x_0)
\rangle_{\rm cum}
\label{cluster}
\end{eqnarray}
with $\langle ~~ \rangle_{\rm cum}$ defined 
in terms of average values over the gauge fields $\langle ~~\rangle$   
and  $F_{\mu\nu}(u,x_0)$ being the path-ordered product of the 
field strength tensor $F_{\mu\nu}(u)$ times two Schwinger strings 
connecting the point $u$ with an arbitrary reference point $x_0$ 
on the surface $S$ appearing in the non-Abelian Stokes theorem.

Equation (\ref{cluster}) is exact. The first cumulant 
vanishes trivially. The second cumulant gives the first non-zero 
contribution to the cluster expansion. In the SVM \cite{DoSi}
one assumes that in the context of heavy quark bound states 
higher cumulants can be neglected and the 
second cumulant dominates the cluster expansion, i. e., 
that the vacuum fluctuations are of a Gaussian type:
\begin{eqnarray}
& & \log \langle W(\Gamma) \rangle  = 
  -{g^2 \over 2} 
\int_S dS_{\mu\nu}(u) \int_S dS_{\lambda\rho} (v)  
\nonumber \\
& & 
\times \langle F_{\mu\nu}(u,x_0)  F_{\lambda\rho}(v,x_0) \rangle_{\rm cum}, 
\label{svm}
\end{eqnarray}
with 
\begin{eqnarray}
& & \langle F_{\mu\nu}(u,x_0)  F_{\lambda\rho}(v,x_0) \rangle_{\rm cum}
\nonumber \\
& &  ={\beta\over g^2} \Bigg\{
(\delta_{\mu\lambda}\delta_{\nu\rho} - 
\delta_{\mu\rho}\delta_{\nu\lambda}) D\left( (u-v)^2 \right) 
\nonumber \\
& &
 + {1\over 2}\left[
{\partial\over\partial u_\mu}\left( (u-v)_\lambda\delta_{\nu\rho}
 - (u-v)_\rho\delta_{\nu\lambda} \right) \right. 
\nonumber \\
& & 
\left.  
+ {\partial\over\partial u_\nu}\left( (u-v)_\rho\delta_{\mu\lambda} 
- (u-v)_\lambda\delta_{\mu\rho} \right) \right]
\nonumber \\
& & \qquad\qquad \times D_1 \left( (u-v)^2 \right) \Bigg\}
\label{cum2}\\
& & 
\beta\equiv{g^2\over 36} 
{\langle {\rm Tr \>} F_{\mu\nu}(0) F_{\mu\nu}(0) 
\rangle \over D(0) + D_1(0)} \> .
\nonumber
\end{eqnarray}
Eqs. (\ref{svm}) and (\ref{cum2}) define the SVM for heavy quarks. 
The correlator functions $D$ and $D_1$ are unknown. 
The perturbative part of $D_1$, which is expected to be dominant 
in the short-range behaviour, can be obtained by means of the 
standard perturbation theory:
\begin{equation}
D_1^{\rm pert} (x^2) = {16 \alpha_{\rm s} \over 3\pi}{1\over x^4} 
+ ~{\rm higher ~~orders} \>. 
\label{D1pert}
\end{equation}
One of the main features of SVM is to insert the information coming 
from the lattice inside the analytic model. Indeed  the non-perturbative 
behaviour of  $D$ and $D_1$ was evaluated on the lattice (see \cite{DoSi})
\begin{equation}
\beta D^{\rm LR}(x^2) =d e^{-\delta|x|},
\> \beta D_1^{\rm LR}(x^2) = d_1 e^{-\delta_1|x|}
\label{D1lr} 
\end{equation}
with  $ \delta = (1 \pm 0.1) \, {\rm GeV}, 
\, d = 0.073 \, {\rm GeV}^4, \, \delta_1 = (1 \pm 0.1) \, {\rm GeV},\, 
d_1 = 0.0254 \, {\rm GeV}^4$.
From Eqs. (\ref{svm})-(\ref{D1lr}) all the potentials follow directly.
In particular the static potential is given by 
\begin{eqnarray}
& & V_0(r) =  \beta \int_{-\infty}^{+\infty} d\tau \left\{ 
\int_0^r d\lambda (r-\lambda)\right.
 \nonumber \\
& & \left.
\times D(\tau^2+\lambda^2)
 + \int_0^r d\lambda {\lambda\over 2}D_1(\tau^2+\lambda^2) \right\},
\label{v0svm}
\end{eqnarray}
In the $r\to 0$ limit (\ref{v0svm}) reproduces one gluon exchange while
in the limit $r\to \infty$ 
\begin{eqnarray}
& & V_0(r)= \sigma_2 r + {1\over 2 } C_2^{(1)} - C_2
\nonumber \\
& & \sigma_2 \equiv \beta \int_{-\infty}^{+\infty} d\tau 
\int_0^\infty d\lambda ~D(\tau^2+\lambda^2) \>, 
\nonumber\\
& & C_2 \equiv \beta \int_{-\infty}^{+\infty} d\tau 
\int_0^\infty d\lambda ~\lambda ~D(\tau^2+\lambda^2), 
\nonumber \\
& & C_2^{(1)} \equiv \beta \int_{-\infty}^{+\infty} d\tau 
\int_0^\infty d\lambda ~\lambda ~D_1(\tau^2+\lambda^2).
\nonumber
\end{eqnarray}
It is clear that (\ref{v0svm}) contains the MAL result for the static 
potential upon identification of the string tension and of the constants.
For the other potentials  see Ref. \cite{BV} (complete form) and Tab. 1 
($r\to \infty$ limit).

\subsection{Dual QCD}

The duality assumption that the long distance physics of a Yang--Mills
theory depending upon strong coupled gauge potentials $A_\mu$ is
the same as the long distance physics of the dual theory describing 
the interactions of weakly coupled dual potentials 
${\cal C}_\mu\equiv \displaystyle \sum_{a=1}^8 C_\mu^a \lambda_a / 2$ 
and monopole fields ${\cal B}_i\equiv \displaystyle \sum_{a=1}^8 B_i^a 
\lambda_a/2$, forms the basis of DQCD \cite{dual}. The model is constructed
as a  concrete realization  of the Mandelstam--t'Hooft  
dual superconductor mechanism of confinement.  Since the main interest 
is solving such a theory in the long-distance regime,
the Lagrangian ${\cal L}_{\rm eff}$ is explicitly constructed  as the minimal
dual gauge invariant extension of a quadratic Lagrangian 
with the further requisite to give a mass to the dual gluons 
(and to the monopole fields) via a spontaneous 
symmetry breaking of the dual gauge group.

We denote by $\langle W_{\rm eff} (\Gamma)\rangle$ the average 
over the fields of the Wilson loop of the dual theory \cite{dual}:
$$
\langle W_{\rm eff} (\Gamma) \rangle= 
{\int {\cal D} C \,{\cal D} B \,
e^{i \int dx [ {\cal L}_{\rm eff} (G_{\mu\nu}^{\rm S}) 
+ {\cal L}_{\rm GF} ] }
\over \int {\cal D} C \,{\cal D} B \, 
e^{i \int dx [ {\cal L}_{\rm eff} (G_{\mu\nu}^{\rm S}=0) 
+ {\cal L}_{\rm GF} ] } }. 
\nonumber
$$ 
${\cal L}_{\rm GF}$ is a gauge fixing term and 
the effective dual Lagrangian in presence of quarks
is given by 
$$ 
{\cal L}_{\rm eff} = 
2~{\rm Tr} 
\left\{ - {1\over 4} {\cal G}^{\mu\nu} {\cal G}_{\mu\nu}
+ {1\over 2} ({\cal D}_\mu {\cal B}_i)^2  \right\} 
- U({\cal B}_i).
\nonumber
$$
$U({\cal B}_i)$ is the Higgs potential with a minimum at a 
non-zero value  ${\cal B}_{01} = B_0 \lambda_7$, 
${\cal B}_{02} = - B_0 \lambda_5$ and ${\cal B}_{03} = B_0 \lambda_2$. 
Moreover we have taken $B_1=B_2=B$, the dual potential proportional 
to the hypercharge matrix ${\cal C}_\mu =C_\mu Y$ and 
\begin{eqnarray}
{\cal D}_\mu {\cal B}_i &=& \partial_\mu {\cal B}_i 
+ i e [ {\cal C}_\mu, {\cal B}_i] \>, \quad\quad  e\equiv{2\pi \over g} \>,
\nonumber \\
{\cal G}_{\mu\nu} &=& \left(
\partial_\mu C_\nu - \partial_\nu C_\mu 
+ G_{\mu\nu}^{\rm S} \right) Y , 
\nonumber \\
G_{\mu\nu}^{\rm S}(x) &\equiv& g
\varepsilon_{\mu\nu\alpha\beta} \int ds \int d\tau 
{\partial y^\alpha \over \partial s} {\partial y^\beta\over \partial \tau}
\nonumber\\ 
& & \qquad\qquad\quad \times \delta(x-y(s,\tau)),
\nonumber
\end{eqnarray}
where $y(s,\tau)$ is a world sheet with boundary $\Gamma$ swept out by the
Dirac string.  

\begin{table*}[hbt]
\setlength{\tabcolsep}{1.5pc}
\newlength{\digitwidth} \settowidth{\digitwidth}{\rm 0}
\catcode`?=\active \def?{\kern\digitwidth}
\caption{Complete MAL potential and long distance SVM and DQCD potentials;
 from parameterization (\ref{D1lr}) we have: $ \sigma_2 
={\pi d\over \delta^2}$,
 $C_2={4 d\over \delta^3}$; $C_2^{(1)}={4 d_1 \over \delta_1^2}$;$D_2= 
{3 \pi d\over\delta^4}$;$ E_2={32 d \over \delta^5}$;
see the values of $d$ and $\delta$ given in the text. The dual gluon mass
 $M^2={\pi\sigma\over  4 \alpha_s}$; typical values are $\sigma=0.18$;
$\alpha_s=0.35$.}
\label{tab:exp}
\begin{tabular*}{\textwidth}{@{}l@{\extracolsep{\fill}}rrrr}
\hline
                 
\cline{1-4} 
                 & \multicolumn{1}{r}{MAL} 
                 & \multicolumn{1}{r}{SVM} 
                 & \multicolumn{1}{r}{DQCD}          \\

\hline
$V_0$ &
$-{4\over 3} {\alpha_s\over r}+ 
\sigma r + C $ & $ \sigma_2 r + {1\over 2} C_2^{(1)} - C_2 $&
$\sigma r -0.646 \sqrt{\sigma \alpha_s}$
\\ & & & \\
$\Delta V_a $ & 0 & {\rm const. self-energy terms} & 
{\rm const. terms}$ - 2{\sigma\over r}$ 
\\ & &  & \\
${dV_1/dr}$ & $-\sigma $ & $-\sigma_2 + {C_2\over r}$ &
$-\sigma + {0.681 \over r} \sqrt{\sigma \alpha_s}$\\
${dV_2/dr}$ & 
${4\over 3}
{\alpha_s\over r^2}$ & ${C_2 \over r}$ & $ {0.681\over r} \sqrt{\sigma
\alpha_s}$ \\
& & & \\
$V_3$ & $ 4 {\alpha_s\over r^3}$ & {\rm exp. fall off} & ${4\over 3}\alpha_s
( M^2 + {3\over r} M + {3\over r^2} ) {e^{-Mr}\over r}$
\\ 
& & & \\
$ V_4$ & $ 32 \pi \alpha_s \delta^3({\bf r})$& {\rm exp.
fall off}& ${4\over 3} \alpha_s M^2 {e^{-Mr}\over r}$ \\
& & & \\
$V_b$ & ${8\over 9} {\alpha_s\over r} 
- {1\over 9} \sigma r$ & $- {1\over 9} \sigma_2 r - {2\over 3}
{ D_2\over r} + {8\over 3} {E_2\over r^2}$ & $ -0.097 \sigma r -0.226 
\sqrt{\sigma \alpha_s}$ \\
& & & \\
$V_c$ & $ -{2\over 3}{\alpha_s\over r}
- {1\over 6} \sigma r$ & $ -{1\over 6}\sigma_2 r - {D_2\over r}
+ {2\over 3} {E_2\over r^2} $& $ -0.146 \sigma r -0.516 \sqrt{\sigma \alpha_s}$
\\
& & & \\
$V_d$ & $ -{1\over 9} \sigma r - {1\over 4} C$ & $ - {1\over 9} \sigma_2 r +
 {1\over 4} C_2 - {1\over 8} C_2^{(1)} + {1\over 3} {D_2 \over r} - {2\over 9}
{E_2\over r^2} $ & $ -0.118 \sigma r +0.275 \sqrt{\sigma\alpha_s}$\\
$V_e$ & $ -{1\over 6} \sigma r$ &$ -{1\over 6} \sigma_2 r + {1\over 2}
{D_2 \over r} - {1\over 3} {E_2\over r^2}$ & $- 0.177 \sigma r + 0.258
 \sqrt{\sigma \alpha_s} $\\
& & &  \\ \hline
\end{tabular*}
\end{table*}

The assumption that the dual theory 
describes the long distance $q\bar q$ interaction in QCD 
then takes the form:
\begin{equation}
\langle W(\Gamma) \rangle  = \langle W_{\rm eff} (\Gamma) \rangle  \>, 
\quad {\rm for ~large ~loops ~\Gamma}.
\label{ass}
\end{equation}
Large loop means that the size $R$ of the loop is
large compared to the inverse mass ($M_H^{-1} \simeq$ (600 MeV)$^{-1}$)
of the Higgs particle (monopole field). Furthermore, since the dual 
theory is weakly coupled at large distances, we can evaluate 
$\langle W_{\rm eff} (\Gamma) \rangle$ via a semiclassical expansion to 
which the classical configuration of the dual potentials and monopoles gives 
the leading contribution. In the leading classical approximation
\begin{equation}
i \log \langle W_{\rm eff}(\Gamma) \rangle 
= - \int dx ~{\cal L}_{\rm eff} (G_{\mu\nu}^{\rm S}) \>,
\end{equation}
with ${\cal L}_{\rm eff} (G_{\mu\nu}^S)$  evaluated
at the solution of the classical equations of motion.
For the static potential we have
\begin{eqnarray}
V_0 &=& - {4\over 3} {\alpha_s\over r} \exp\left(-0.511 \sqrt{\sigma\over
\alpha_s}r\right) 
\nonumber\\
&+& \sigma r - 0.646 \sqrt{\sigma \alpha_s}
\nonumber
\end{eqnarray}
This results directly from the classical solution to the nonlinear equations 
obtained from ${\cal L}_{\rm eff}$
and reproduces the one gluon exchange in the limit $r\to 0$ and the 
string term in the limit $r \to \infty$. The coefficient of the exponent
can be actually given in terms of the dual gluon mass $M=6 g^2 B_0^2
\simeq {\pi\over 4} {\sigma \over \alpha_s}$.
An interpolation of the numerical results for the potentials 
can be found in \cite{BV,dual}. In Tab.1 we report the potentials in 
$ r\to \infty$ limit.

\section{FLUX TUBE STRUCTURE AND CON\-FI\-NE\-MENT ME\-CHA\-NISM}

To dis\-cuss the 
se\-mi\-re\-la\-ti\-vi\-stic non-per\-tur\-ba\-ti\-ve dynamics is convenient 
to study the large $r$ limit in the three above models as reported in Tab.1.
The MAL result is the realization of the intuitive Buchm\"uller's picture of 
zero magnetic field in the flux tube comoving system: ${dV_2/dr}$ which is
given by the magnetic interaction is zero in the non-perturbative region;
similarly the velocity dependent potentials $V_{\rm b}$--$V_{\rm e}$ 
come from the consideration of the flux--tube energy. From Tab.1 it is 
apparent that at the leading order in the long--range limit, neglecting 
exponentially falling off terms, the SVM contains exactly the MAL model 
results. In the spin dependent sector of the potential, 
both the SVM and DQCD not only reproduce the long-range behaviour  
given by the area law, but also give $1/r$ corrections  
to $dV_1/dr$ and $dV_2/dr$. These corrections are equal 
in both models and very near to the absolute value of the 
constant term in the static potential (the SVM also supplies
for the explication of this fact). This perfect agreement is 
absolutely not trivial and seems to be very meaningful, since it arises  
from two very different models in a region of distances 
in which the physics cannot be described by the area law alone.
It is  now clear that the vanishing of the magnetic part    
in the non-perturbative region takes place only at the leading order  
in the long-range limit. Therefore, working in a Bethe--Salpeter context, 
there is no need to assume an effective pure convolution kernel 
which is a Lorentz scalar. Velocity dependent contributions to 
the quark-antiquark potential are important. In fact the string behaviour 
of the non-perturbative interaction shows up when we consider 
the velocity dependent part of the potential and this is also what the data 
require. The velocity dependent structure which arises from the DQCD model 
differs slightly in the coefficients with respect to the area law behaviour. 
In conclusion SVM and DQCD reproduce the flux tube  
distribution measured on the lattice and the general features coming from 
the area law. Both give analytical expressions for the Wilson loop  which 
describe the evolving behaviour of $\langle W(\Gamma)\rangle$ 
from the short to the long distances but not all predictions are equal 
in the two models in the intermediate distances region, 
in particular in the velocity dependent sector of the potential, and also 
in the spin-spin interaction as well as in $\Delta V_a$. The up to now 
available lattice data \cite{bali} confirm the MAL results but are not 
sufficiently accurate to discriminate between SVM and DQCD models.

\end{document}